\documentstyle[11pt,mynewpasp,twoside,epsf]{article}
\def\edcomment#1{\iffalse\marginpar{\raggedright\sl#1\/}\else\relax\fi}
\reversemarginpar

\begin{document}
\title{Early-type Galaxy Halo Dynamics inferred using \\ 
 the PN Spectrograph}
 \author{N.G. Douglas$^1$, A.J. Romanowsky$^2$,
    K. Kuijken$^{1,3}$, M.R. Merrifield$^2$,\\
    N.R. Napolitano$^1$, M.  Arnaboldi$^4$,
    K.C.  Freeman$^5$, M. Capaccioli$^6$,\\ 
    O. Gerhard$^7$ }

\affil{$^1$Kapteyn Institute, Groningen; 
       $^2$School of Physics \& Astronomy, University of 
		Nottingham; 
	$^3$University of Leiden;
	$^4$INAF-Osservatorio Astronomico di 
		Pino Torinese, Turin;
	$^5$RSAA, Mt. Stromlo and Siding Spring Observatories;
	$^6$INAF-Osservatorio 
		Astronomico di Capodimonte, Naples;
	$^7$Astronomisches Institut, University of Basel}

\begin{abstract} 
 A new instrument is providing crucial data with which to probe
the structure of dark halos in elliptical galaxies
 \end{abstract}

\section{Introduction}

 Although evidence for dark halos in spiral galaxies is well-founded on
kinematic measurements, similar studies in elliptical galaxies have not
been so conclusive.  One difficulty has been the lack of a suitable
kinematic tracer at radii beyond the point at which integrated stellar
spectroscopy becomes impractical.  In the late 1980s the situation
improved when planetary nebulae (PNe) were recognised as a powerful
diagnostic, and since 2001 our team has been operating an instrument,
the Planetary Nebula Spectrograph, which is
specially designed for these observations. This results
in a leap forward in
detection efficiency and makes possible a range of new projects.

\section{The PN.Spectrograph}

The PN.S incorporates two highly efficient slitless spectrographs
back-to-back.  Via a common [OIII] filter this results in simultaneous
images of the field in which the PNe are easily distinguished from
stars.  Their velocities can be derived from the relative shifts in
their positions in the two images.  This ``Counter-Dispersed Imaging"
technique is described in Douglas et al.\ (2002), along with other details
of the project. Its value lies in the high optical efficiencies attained, in
the absence of any astrometric requirements, and in the fact that it 
is a single-step procedure.

\begin{figure}
\plottwo{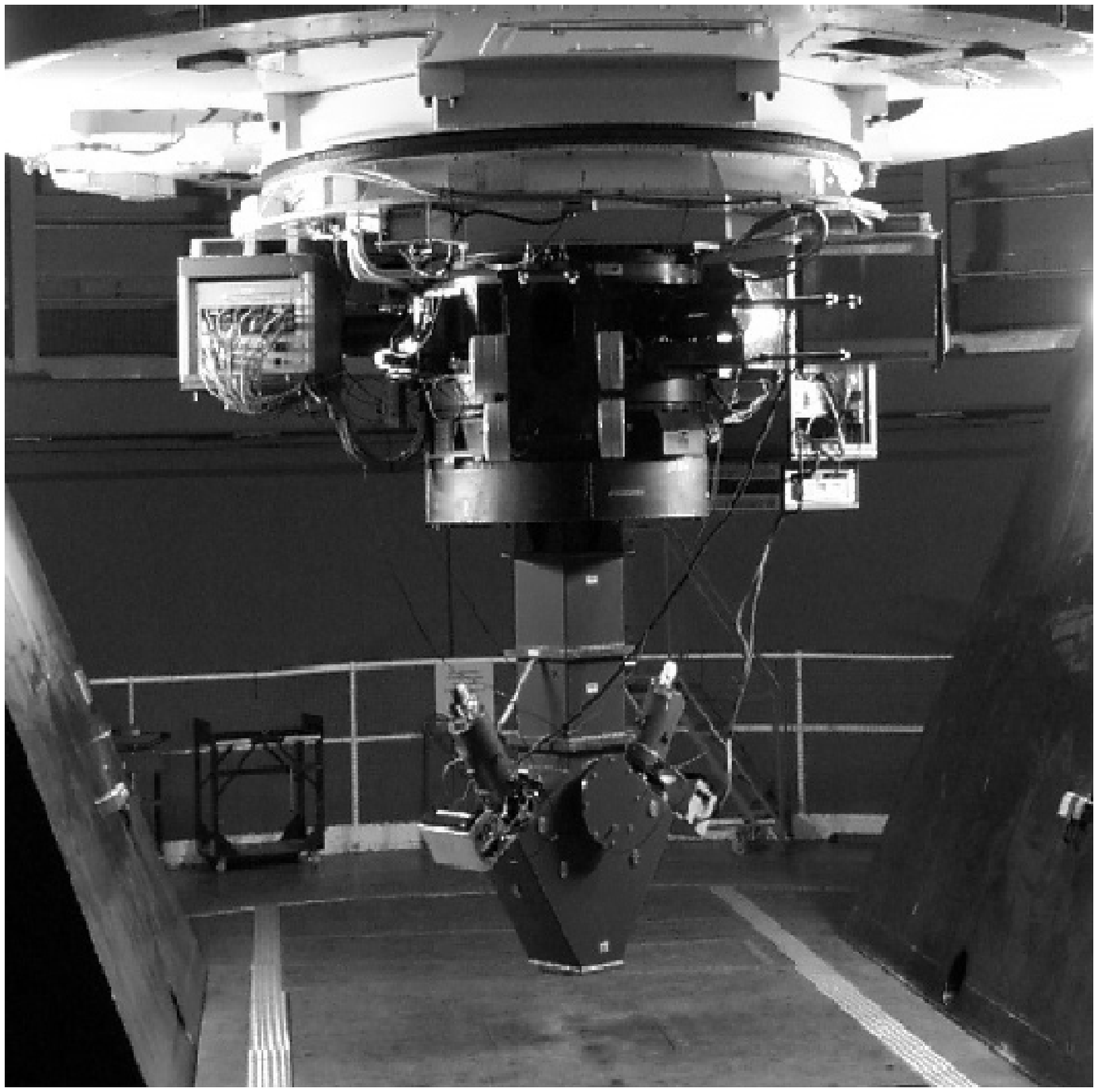}	
          {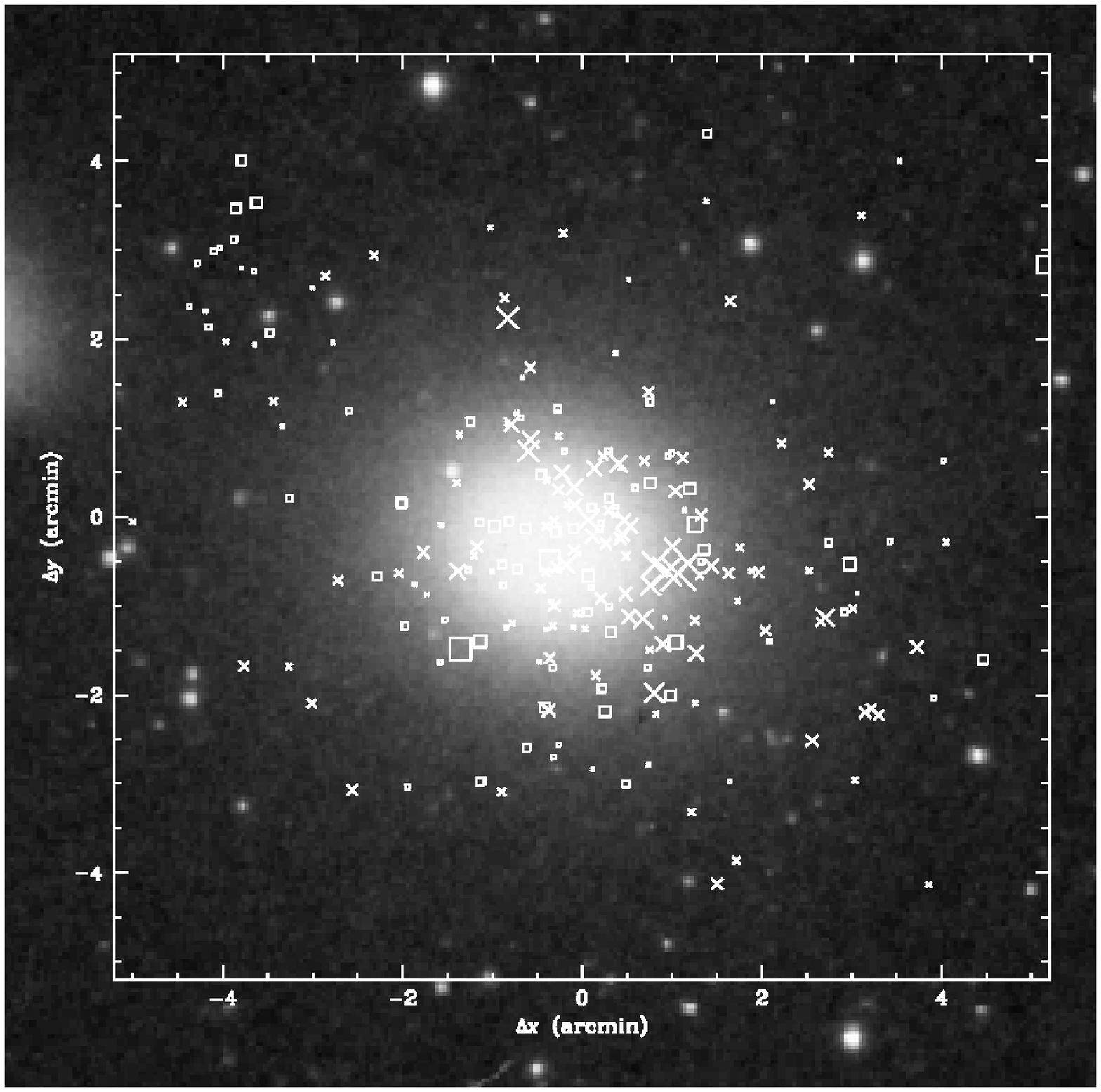}  
\caption{Left: the PN.S at the Cassegrain focal station of the
William Herschel Telescope, which has built-in spectral
lamps for calibrating the measured velocities. The PNS measures
CDI images simultaneously by means of the two detectors 
shown here (photo courtesy of R.A. Hijmering).  
Right: the elliptical galaxy NGC~3379 with 230 PNe
superposed,  the symbols indicating magnitude and sign of the
velocity.
}

\end{figure}

Our first project, currently in progress, 
is to examine the dark matter halos of
apparently round, moderate-luminosity elliptical galaxies, too
small to have been easily addressed by X-ray or gravitational-lensing
analysis. Like most ellipticals, they do not produce much emission
in neutral hydrogen either. Finally, conventional stellar spectroscopy
does not probe far enough out, where dark-matter is expected to dominate.

In the prevailing CDM model of galaxy formation the mass distribution
is such that, over the region probed by our measurements, galaxies
are expected to have a flat or slightly rising circular velocity profile
(Navarro, Frenk, \& White 1996). 
This is  in agreement
with the asymptotically flat rotation curves of spiral galaxies and has
also been  borne out for the larger ellipticals
(Gerhard et al.\ 2001). 
However the first three objects in
our sample, NGC~3379, NGC~821 and NGC~4494, yielded the surprising
result (though seen earlier in the case of
NGC~3379 by Ciardullo, Jacoby \& Dejonghe 1993) 
that the velocity dispersion strongly declines with radius.
With realistic assumptions about the isotropy of the
velocity distribution  
this is indicative of a falling circular velocity profile and is
almost consistent with there being no dark halo at all -- see Romanowsky
et al.\  elsewhere in these proceedings, and Romanowsky et al.\ (2003).

\end{document}